\documentclass[fleqn,twoside]{article}
\usepackage{amssymb}
\usepackage{espcrc2}

\usepackage{graphicx}

\newcommand{\AmS}{{\protect\the\textfont2
  A\kern-.1667em\lower.5ex\hbox{M}\kern-.125emS}}

\title{Magnetic order and moment distribution in doped spin-chain systems}

\author{Sebastian Eggert\address[CH]{Physics Dept., Univ. of Kaiserslautern, 67663 Kaiserslautern, Germany}
\thanks{On leave from
Institute of Theoretical Physics, Chalmers
University of Technology,
S-412 96 G\"oteborg, Sweden},
Ian Affleck\address{ Department of Physics, University of British Columbia,
Vancouver, BC, Canada V6T1Z1}}
       
\begin{document}

\begin{abstract}
We consider the effect of doping on the magnetic order in 
quasi one-dimensional (1D) 
antiferromagnets.   First the detailed magnetic response 
of finite spin-1/2 chains in the presence of a staggered field is
determined, which 
by itself shows an interesting crossover as function of length and 
 field strength.  We can then understand the ordering by including an
effective coupling between the chains, which results in
a N\'eel ordered phase at low temperatures, but with a 
broad distribution of
magnetic moments  that has to be determined self-con\-sistently.\\~\\
{\it PACS numbers}: 75.10.Jm, 75.20.Hr \hfill  
\end{abstract}

\maketitle

Doping and impurity effects in low-dimensional antiferromagnets
remain of strong interest in the condensed
matter physics community, spurred by high-$T_c$ superconductivity
and other exotic effects in those systems.  Typically, antiferromagnetic 
three dimensional N\'eel order d\"{o}s exist at extremely low temperatures,
but is easily destroyed in such systems by a combination of  quantum 
fluctuations, temperature fluctuations, and disorder effects.  
In this paper, we study how static impurities affect the 
N\'eel order in quasi-1D antiferromagnets, which are modelled by spin-1/2
chains with a weak coupling $J'\ll J$ to the neighboring 
chains $\vec \delta$ (before doping)
$H =  \sum_{j,\vec{y}} \left(J \vec{S}_{j,\vec{y}} \cdot \vec{S}_{j+1,\vec{y}}
+ \sum_{\vec{\delta}} J' \vec{S}_{j,\vec{y}} \cdot \vec{S}_{j,\vec{y}+
\vec{\delta}}\right).$

In uncoupled chains it is known that long-range antiferromagnetic order 
is destroyed by quantum fluctuations even at $T=0$, which results
in antiferromagnetic correlations that decay with a powerlaw.  
However,
a staggered field $B$ can easily induce alternating order 
$m_{\rm alt} = \frac{1}{L}\sum_j (-1)^j \langle S^z_j\rangle$ 
throughout the sample
since the staggered susceptibility diverges at $T=0$ \cite{imry}. 

In the case of quasi-1D antiferromagnets the coupling to the neighboring
chains plays the role of an effective 
staggered field from the antiferromagnetic background
$B = z J'  m_{\rm alt}$, which always implies an
ordered state at $T=0$ \cite{imry}.  
Here, $z$ is the number of neighboring chains. 
Note, that the effective field 
neglects quantum fluctuations between neighboring chains, but within
each chain we are able to determine the alternating moment by 
a full quantum mechanical solution $m_{\rm alt}(B)$.  
We can then study the ordered state by solving the effective 
field equation self-consistently for a non-zero solution $m$
\begin{equation}
m = m_{\rm alt}(z J' m). \label{m}
\end{equation}
For the undoped case it
is known that the alternating magnetization is given by a scaling law
$m_{\rm alt}(B) =a B^{1/3}$ up to logarithmic corrections \cite{affleck1},
which results in $m= a^{2/3} \sqrt{zJ'}$ \cite{affleck2}
in good agreement with experiments\cite{kojima}.  

In a doped system the impurities effectively 
cut the chain at low temperatures\cite{PRB92}.  It is
therefore necessary to calculate the alternating moment 
for finite chains and average over the disorder in order to arrive
at a self-consistent distribution of moments $m$ in Eq.~(\ref{m}).
Previous results have shown that finite chains have a
different staggered susceptibility as a function of $T$ 
depending if they contain an even or odd number 
of spins\cite{PRL02}, which can be traced to the contribution of the zero-modes
in the correlation functions\cite{mattsson}.   
Moreover, it has been found that
the response is lower for finite chains which yields a distinct 
drop of the the N\'eel temperature with increased  doping \cite{PRL02} in 
agreement with existing experiments \cite{kojima}.  
The undoped value of $T_N$
scales roughly with $J'$ \cite{imry} up to logarithmic corrections which we
have quantitatively determined to give $ 0.6 \lesssim T_N/zJ' \lesssim 1$  
in the range $0.1>J'/J>0.00001$.

\begin{figure}
\includegraphics[width=0.45\textwidth]{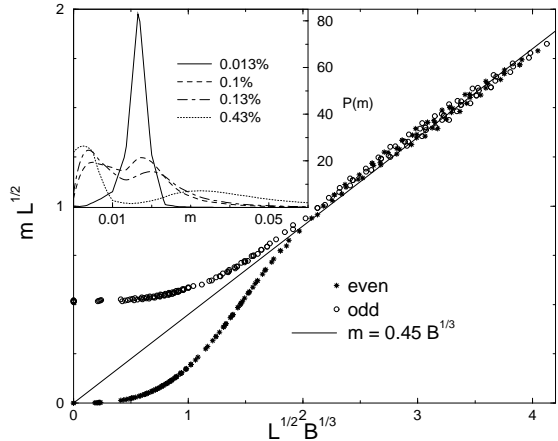}
\caption{The alternating magnetization as a function of staggered
magnetic field $B$
and length $L$ from numerics for open chains with
$L\leq27$ and $B\leq0.5J$. Inset: The corresponding 
distribution of moments in units of 
$\hbar g \mu_B$ according to Eq.~(\ref{m})
for $J'=0.0008J$.
The non-uniform distribution of $m$ along the chains
was taken into account by including the known behavior near the edges 
\cite{PRL02,PRL95}.} \label{malt}
\end{figure}
The alternating moment in a finite field at $T=0$ in Fig.~1 
shows again a distinct difference
between even and odd chains and obeys a scaling behavior of the form
\begin{equation}
m_{\rm alt}(B) = L^{-1/2}f\left(L^{1/2}B^{1/3}\right),\label{scaling}
\end{equation}
in agreement with a dimensional scaling analysis.
Assuming a sharp distribution of lengths $L\approx 1/p$ Eq.(\ref{scaling}) 
also implies a scaling form for the average alternating moment
\begin{equation}
m = L^{-1/2} g(zJ'L).
\end{equation}
The low doping limit $\lim_{x\to\infty} f(x) \to a x$ 
recovers the $B^{1/3}$ scaling
behavior with a length-independent
solution for $m = a^{2/3} \sqrt{zJ'}$, which we 
see as a sharp peak in the moment distribution for low doping levels in
the inset of Fig.~\ref{malt}. For larger doping $p$ we 
expect a broader distribution, but 
the difference between even and odd chains also becomes important.
The self-consistent
solution to Eq.~(\ref{m}) shows  two distinct 
peaks corresponding to the even and odd chain segments as displayed in 
the inset of Fig.~\ref{malt}.  Interestingly,
the peak for the odd chains shifts to higher moments 
(increased order parameter)
as the doping increases, even though the N\'eel temperature is known
to drop.  In fact we know from Fig.~\ref{malt} that 
$m_{\rm alt} \propto L^{-1/2}$ diverges for short odd chains. 
 However, in that limit 
quantum fluctuations between the shorter chains also become 
more important so it is not clear if this exotic 
``increased order from disorder'' effect is robust or if
the order parameter always decreases with increasing doping. 
Qualitatively, however, 
the peak will indeed split for larger doping $p \gtrsim z J'/10$ 
and  the shape
of the distributions in the inset is adequate, except for a 
doping dependent rescaling of $m$.

In summary, we have studied the doping dependence of the 
N\'eel ordered phase at $T=0$ in quasi-1D antiferromagnets and showed
explicitly how the order parameter is modified in the presence of
static disorder.
The distribution of the magnitude of alternating moments broadens 
with doping and we find a characteristic double peak structure for
$p \gtrsim z J'/10$.

\end{document}